\documentclass{amsart}
\usepackage{amssymb}
\usepackage{bbm} 
\usepackage{graphicx}
\usepackage{color}
\graphicspath{{Figures/}}
\usepackage[hidelinks]{hyperref}
\newtheorem{theorem}{Theorem}[section]
\newtheorem{lemma}[theorem]{Lemma}
\theoremstyle{remark}

\numberwithin{equation}{section}

\def\f{\mathbf{f}}
\def\g{\mathbf{g}}
\def\w{\mathbf{w}}
\def\x{\mathbf{x}}
\def\m{\boldsymbol{\mu}}
\def\RR{\mathbb{R}}
\begin{document}
\title[Statistical and numerical considerations of Backus average]{Statistical and numerical considerations of Backus-average product approximation}
\author{Len Bos}
\address{Dipartimento di Informatica, Universit\`a di Verona, Italy}
\email{leonardpeter.bos@univr.it} 
\author{Tomasz Danek}
\address{Department of Geoinformatics and Applied Computer Science, AGH--University of Science and Technology, Krak\'ow, Poland}
\email{tdanek@agh.edu.pl}
\author{Michael A. Slawinski}
\address{Department of Earth Sciences, Memorial University of Newfoundland, St. John's, Newfoundland, Canada}
\email{mslawins@mac.com}
\author{Theodore Stanoev}
\address{Department of Earth Sciences, Memorial University of Newfoundland, St. John's, Newfoundland, Canada}
\email{theodore.stanoev@gmail.com}

\subjclass[2000]{74B05, 86A15, 86-08}

\date{December 14, 2018}

\dedicatory{\begin{flushleft}
This version contains corrections of typographical errors in Bos, L., Danek, T., Slawinski, M.A., Stanoev, T. (2018) Statistical and numerical considerations of Backus-average product approximation. Journal of Elasticity {\bf 132}(1), 141--159.
\end{flushleft}}

\keywords{Backus averaging, Continuum mechanics, Approximation, Statistical analysis, Numerical analysis}
\begin{abstract}
In this paper, we examine the applicability of the approximation, $\overline{f\,g}\approx \overline f\,\overline g$\,, within Backus~\cite{Backus1962} averaging.
This approximation is a crucial step in the method proposed by Backus~\cite{Backus1962}, which is widely used in studying wave propagation in layered Hookean solids.
According to this approximation, the average of the product of a rapidly varying function and a slowly varying function is approximately equal to the product of the averages of those two functions.

Considering that the rapidly varying function represents the mechanical properties of layers, we express it as a step function.
The slowly varying function is continuous, since it represents the components of the stress or strain tensors.
In this paper, beyond the upper bound of the error for that approximation, which is formulated by Bos et al.~\cite{BosEtAl2017}, we provide a statistical analysis of the approximation by allowing the function values to be sampled from general distributions.

Even though, according to the upper bound, Backus~\cite{Backus1962} averaging might not appear as a viable approach, we show that---for cases representative of physical scenarios modelled by such an averaging---the approximation is typically quite good.
We identify the cases for which there can be a deterioration in its efficacy.

In particular, we examine a special case for which the approximation results in spurious values.
However, such a case---though physically realizable---is not likely to appear in seismology, where Backus~\cite{Backus1962} averaging is commonly used.
Yet, such values might occur in material sciences, in general, for which Backus~\cite{Backus1962} averaging is also considered.
\end{abstract}
\maketitle
\section{Introduction}
Let us consider a Hookean solid, which is expressed by fourth-rank tensors in accordance with Hooke's law,
\begin{equation}
\label{eq:Hooke}
\sigma_{ij}
=\sum\limits_{k=1}^3\sum_{\ell=1}^3c_{ijk\ell}\,\varepsilon_{k\ell}\,,
\qquad i,j\in\{1,2,3\} 
\,,
\end{equation}
which relates the stress,~$\sigma$\,, and strain,~$\varepsilon$\,, tensors.
Backus~\cite{Backus1962} showed that a homogeneous transversely isotropic Hookean solid can be long-wave equivalent to a stack of thin isotropic or transversely isotropic layers.
Bos et al.~\cite{BosEtAl2017} examined the mathematical underpinnings of the Backus~\cite{Backus1962} approach, in the context of generally anisotropic layers.
Readers interested in an overview, a motivation or details of equivalent media might refer to these papers or to Slawinski~\cite[Section~4.2]{SlawinskiWS2016}.
However, there remains an examination of the underlying assumption.
Hence, this paper.

Backus~\cite{Backus1962} writes

\begin{quote}
The only approximation that we make in the present paper is the following: if
$f(x_3)$ is nearly constant when $x_3$ changes by no more than $\ell'$\,, while
$g(x_3)$ may vary by a large fraction over this distance, then, approximately,
$\overline{f\,g}\approx \overline f\,\overline g$\,.
\end{quote}

\noindent In our presentation, for conciseness of notation, $x$ stands for $x_3$\,.

Following the definition proposed by Backus~\cite{Backus1962}, the average of the function $f(x)$ of ``width''~$\ell'$ is the moving average given by
\begin{equation}
\label{eq:BackusOne}
\overline f(x)
:=\int\limits_{-\infty}^\infty w(\zeta-x)f(\zeta)\,{\rm d}\zeta \,,
\end{equation}
where the weight function,~$w(x)$\,, has the following properties:
\begin{equation*}
w(x)\geqslant0\,,\!\!
\quad w(\pm\infty)=0\,,
\int\limits_{-\infty}^\infty w(x)\,{\rm d}x=1\,,
\int\limits_{-\infty}^\infty x\,w(x)\,{\rm d}x=0\,,
\int\limits_{-\infty}^\infty x^2 w(x)\,{\rm d}x=(\ell')^2
\,.
\end{equation*}
Within that context, Bos et al.\ \cite[Lemma 3]{BosEtAl2017} prove the following lemma, which may be restated as follows.
\begin{lemma}
\label{lem:Len}
Given $f(x)$ that is nearly constant along an interval of length~$\ell'$\,, and $g(x)$\,, which is allowed to vary
by a large amount over this interval, we can use the following approximation:
\begin{equation}
\label{eq:Lemma1}
\overline{f\,g}\approx\overline f\,\overline g\,,
\end{equation}
where an overline, $\overline{\,\,\left.\right.\,\,}$\,, denotes an average.
\end{lemma}

\noindent Also, Bos et al.\ \cite{BosEtAl2017} present an upper bound for the error of the approximation in question.
If $f(x)$ is continuous and $g(x)\geqslant 0$\,, then, by the Mean-value Theorem for Integrals,
\begin{equation*}
\overline{f\,g}
=\int\limits_{-\infty}^\infty\!\!f(x)\,g(x)\,W(x)\,{\rm d}x
=f(c)\!\!\int\limits_{-\infty}^\infty\!\!g(x)\,W(x)\,{\rm d}x
=f(c)\,\overline g
\,,
\end{equation*}
for some $c$\,, where for a fixed $x$\,, we set $W(\zeta):=w(\zeta-x)$\,. 
Hence,
\begin{equation*}
\overline{f\,g}-\overline f\,\overline g
=f(c)\,\overline g-\overline f\,\overline g
=\left(f(c)-\overline f\,\right)\overline g\,.
\end{equation*}
As shown explicitly in~\ref{app:Len}, this implies that
\begin{equation}
\label{eq:Len}
\left|\,\overline{f\,g}-\overline f\,\overline g\,\right|
=\left|\,f(c)-\overline f\,\right|\overline g
\leqslant \|f'\|_\infty\left(\,\int\limits_{-\infty}^\infty\left|c-\zeta\right|\,W(\zeta)\,{\rm d}\zeta\right)\overline g
\,.
\end{equation}
If $\|f'\|_\infty$ and $\overline g$ are not exceedingly large and the weight function is reasonable, the {\it absolute} difference between the average of the product and the product of the averages is small. Sometimes, however, it is more useful to measure the {\it relative} error defined as
\begin{equation*}
\frac{\overline{f\,g}-\overline f\,\overline g}{\overline{f\,g}}\times 100\%
\,.	
\end{equation*}
If $\overline{g}=0$\,, this error becomes $100\%$\,; hence, the case of $\overline{g}=0$ is of concern, and we discuss it in Section~\ref{sub:Hooke}.

To obtain expression~(\ref{eq:Len}), for a fixed value of $x$\,, we set $W(\zeta):=w(\zeta-x)$\,, as discussed by Bos et al.~\cite[Appendix~C]{BosEtAl2017}.
Then, $W\geqslant 0$ and
$\int_{-\infty}^\infty W(\zeta)\,{\rm d}\zeta =1$\,.
With this notation, equation~(\ref{eq:BackusOne}) becomes
\begin{equation*}
\overline f
:=\int\limits_{-\infty}^\infty\!\!f(x)\,W(x)\,{\rm d}x
\,.
\end{equation*}
Similarly,
\begin{equation*}
\overline g
:=\int\limits_{-\infty}^\infty\!\!g(x)\,W(x)\,{\rm d}x
\qquad
{\rm and}
\qquad
\overline{f\,g}
:=\int\limits_{-\infty}^\infty\!\!f(x)\,g(x)\,W(x)\,{\rm d}x\,.
\end{equation*}

The purpose of this paper is to use statistical analysis to gain an insight into implications of Lemma~\ref{lem:Len} in both theoretical and pragmatic considerations.
In particular, we examine approximation~(\ref{eq:Lemma1}), namely, $\overline{f\,g}\approx \overline f\,\overline g$\,, which is necessary for the Backus~\cite{Backus1962} averaging process.

In accordance with Backus~\cite{Backus1962} and Bos et al.~\cite{BosEtAl2017}, we associate~$g$ with the elasticity parameters,~$c_{ijk\ell}$\,, contained in expression~(\ref{eq:Hooke}); these values can change abruptly from layer to layer.
For a stack of parallel layers along the $x_3$-axis, we associate the slowly varying function,~$f$\,, as components of the strain tensor,~$\varepsilon_{11}$\,, $\varepsilon_{12}$\,, $\varepsilon_{22}$\,, or the stress tensor,~$\sigma_{i3}$\,, where $i=1,2,3$\,.
These components are constant for the static case and---for a far-field wave propagation---are assumed to be nearly so along the $x_3$-axis, which is normal to the parallel layers.

We begin this paper by formulating the statistical approach to study Lemma~\ref{lem:Len}.
Then, we proceed to numerical examination of several cases of particular pertinence for this study.
We conclude this paper by discussing the wide range of validity of the approximation given in expression~(\ref{eq:Lemma1}), and the single case of its failure.
\section{Statistical approach}
\subsection{General formulation}
To examine the approximation in expression~(\ref{eq:Lemma1}), we consider a medium composed of $n$ parallel layers whose thicknesses vary.
Herein, $f(x)$ is continuous on $[0,L]$ and $g(x)$ is a step function on the same closed interval with breaks at $0=x_0<x_1<\cdots < x_n=L$\,, thus delineating $n$ layers extending to depth~$L$\,.

Let $g_k$ be the value of $g(x)$ on the $k$th interval, $[x_{k-1},x_k]$\,, where $1\leqslant k\leqslant n$\,.
Hence, the average of the product is
\begin{align*}
\overline{f(x)\,g(x)} 
&:= \frac{1}{L}\int\limits_0^L f(x)\,g(x)\,{\rm d}x
=\frac{1}{L}\sum_{k=1}^n\,\int\limits_{x_{k-1}}^{x_k}f(x)\,g(x)\,{\rm d}x
=\frac{1}{L}\sum_{k=1}^n g_k\int\limits_{x_{k-1}}^{x_k}\!\!f(x)\,{\rm d}x\\
&=\frac{1}{L}\sum_{k=1}^n (x_k-x_{k-1})\left\{\frac{1}{x_k-x_{k-1}}\int\limits_{x_{k-1}}^{x_k}\!\!f(x)\,{\rm d}x\right\}g_k
=\sum_{k=1}^n w_k\,f_k\,g_k\,,
\end{align*}
where $w_k:=(x_k-x_{k-1})/L$ is the fraction of the depth of the $k$th layer with respect to the total depth, and
\begin{equation*}
f_k:=\frac{1}{x_k-x_{k-1}}\int\limits_{x_{k-1}}^{x_k}\!\!f(x)\,{\rm d}x
\end{equation*}
is the average of $f(x)$ over the $k$th layer.
Similarly,
\begin{equation*}
\overline{f}
=\frac{1}{L}\int\limits_0^Lf(x)\,{\rm d}x
=\sum_{k=1}^n w_k\,f_k
\qquad{\rm and}\qquad
\overline{g}=\sum_{k=1}^n w_k\,g_k
\,.
\end{equation*}
Herein, the weights are such that $w_k\ge0$ and $\sum_{k=1}^nw_k=1$\,.
Thus, the averages under consideration,  namely, $\overline{f(x)}$\,, $\overline{g(x)}$ and $\overline{f(x)\,g(x)}$\,, are but discrete weighted averages involving three vectors, $\f\in\RR^n$\,, $\g\in\RR^n$ and $\f\,\g\in\RR^{n}$\,, whose components are $f_k$\,, $g_k$ and  $f_k\,g_k$\,, respectively.

In this context, the difference between the average of the product and the product of the averages is
\begin{equation}
\label{eq:E(f,g)}
E(\f,\g)
:=\overline {\f\,\g} -\overline{\f}\,\overline{\g}
\,,	
\end{equation}
where, for any vector $\x\in\RR^n$\,, we set
\begin{equation*}
\overline{\x}
:=\sum_{k=1}^n w_k\,x_k
\,.
\end{equation*}
It is convenient to express $E(\f,\g)$ in matrix-vector form.
\begin{lemma}\label{Efg}
Suppose that $\w\in\RR^n$ is the vector of weights $w_k$\,, and that $W\in\RR^{n\times n}$ is the diagonal matrix with $W_{kk}=w_k$\,.
Then
\begin{equation}
\label{eq:Efg}
E(\f,\g)=\f^tQ\,\g\,,
\end{equation}
where $Q:= W-\w\,\w^t\in\RR^{n\times n}$\,.
\end{lemma}
\noindent {\bf Proof.} It suffices to note that
\begin{align*}
E(\f,\g)
&=\sum_{k=1}^n w_k\,f_k\,g_k-\left(\sum_{k=1}^n w_k\,f_k\right)
\left(\sum_{k=1}^n w_k\,g_k\right)\\
&=\f^t W\,\g - (\f^t \w)(\w^t \g)
=\f^t W\,\g - \f^t(\w\,\w^t)\,\g
=\f^t(W-\w\,\w^t)\,\g\,,
\end{align*}
which is the required result.\,$\square$\\

\medskip \noindent {\bf Remark.} Since $Q\in\RR^{n\times n}$ is symmetric, $E(\f,\g)$ is but a certain bilinear form.
In this discrete case there is a simple, but useful, upper bound for $|E(\f,\g)|.$

\begin{lemma}\label{Ebnd} 
We have
\begin{equation*}
|E(\f,\g)|
\leqslant \Bigl\{\overline{(\f-\overline{\f}\,)^2}\Bigr\}^{1/2}\Bigl\{\overline{\g^2}\Bigr\}^{1/2}
\,.
\end{equation*}
\end{lemma}
\noindent {\bf Proof.} We express
\begin{align*}
E(\f,\g)
&=\sum_{k=1}^n w_k\,f_k\,g_k-\Bigl(\sum_{k=1}^n w_k\,f_k\Bigr)
\Bigl(\sum_{k=1}^n w_k\,g_k\Bigr)\\
&=\sum_{k=1}^n w_k\,f_k\,g_k-\overline{\f}\,\Bigl(\sum_{k=1}^n w_k\,g_k\Bigr)
=\sum_{k=1}^n w_k\,(f_k-\overline{\f}\,)\,g_k
\,.
\end{align*}
Hence, by the weighted Cauchy-Schwartz inequality,
\begin{equation*}
|E(\f,\g)|\leqslant 
\Bigl\{\sum_{k=1}^n w_k\,(f_k-\overline{\f}\,)^2\Bigr\}^{1/2}
\Bigl\{\sum_{k=1}^n w_k\,g_k^2\Bigl\}^{1/2}
=\Bigl\{\overline{(\f-\overline{\f}\,)^2}\Bigr\}^{1/2}
\Bigl\{\overline{\g^2}\Bigr\}^{1/2}
\,,
\end{equation*} 
which is the required result.\,$\square$\\

\medskip\noindent We note that this bound is sharp in the sense that it is attained precisely if~$\g=c\,(\f-\overline{\f}\,)$ for some constant~$c$\,.

Besides giving upper bounds for $|E(\f,\g)|$\,, we may also perform a statistical analysis.
Specifically, suppose that $\f\in\RR^n$ is a random variable sampled from a distribution whose mean is $\boldsymbol{\mu}_f\in\RR^n$ and whose covariance matrix is $C_f\in\RR^{n\times n}$\,.
The correlation matrix is
\begin{equation*}
(C_f)_{ij}=\mathbb{E}((f_i-(\boldsymbol{\mu}_f)_i)(f_j-(\boldsymbol{\mu}_f)_j))\,,\qquad 1\leqslant i,j\leqslant n
\,,
\end{equation*}
which in matrix form becomes
\begin{equation*}
C_f=\mathbb{E}((\f-\boldsymbol{\mu}_f)(\f-\boldsymbol{\mu}_f)^t)
\,;
\end{equation*}
herein, $\mathbb{E}(\cdot)$ refers to the mean of the random variable.
Note that the diagonal entries, 
\begin{equation*}
(C_f)_{ii}=\mathbb{E}((f_i-(\mu_f)_i)^2)
\,,
\end{equation*}
are the variances of the components~$f_i$\,.
Also note that, if the components of $\f$ are independent of each other,~$C_f$ is a diagonal matrix. 

Similarly, we suppose that $\g\in\RR^n$ is a random variable sampled from a distribution whose mean is $\boldsymbol{\mu}_g\in\RR^n$ and whose covariance matrix is $C_g\in\RR^{n\times n}$\,.
Furthermore, it is important to suppose that $\f$ and $\g$ are {\it independent} of one another.

With these assumptions, we may compute the mean and variance of our error statistic,~$E(\f,\g)$\,, which is given in expression~(\ref{eq:Efg}).
\begin{lemma}
\label{Estats}
We have
\begin{equation*}
\mathbb{E}(E(\f,\g))
=(\boldsymbol{\mu}_f)^tQ\,(\boldsymbol{\mu}_g)
=E(\boldsymbol{\mu}_f,\boldsymbol{\mu}_g)
\,,
\end{equation*}
\begin{equation*}
\mathbb{E}((E(\f,\g))^2)
={\rm tr\!}\left[Q\,\mathbb{E}(\f\,\f^t)\,Q\,\mathbb{E}(\g\,\g^t)\right]
={\rm tr\!}\left[Q\,(C_f+\boldsymbol{\mu}_f(\boldsymbol{\mu}_f)^t)\,Q\,
(C_g+\boldsymbol{\mu}_g(\boldsymbol{\mu}_g)^t)\right]
\end{equation*}
and
\begin{equation*}
{\rm var\!}\left[E(\f,\g)\right]
={\rm tr\!}\left[Q\,C_f\,Q\,C_g+Q(\boldsymbol{\mu}_f(\boldsymbol{\mu}_f)^t)\,Q\,C_g+Q\,C_f\,Q(\boldsymbol{\mu}_g(\boldsymbol{\mu}_g)^t)\right]
\,.
\end{equation*}
\end{lemma}
\noindent {\bf Proof.} For the mean, we compute
\begin{equation*}
\mathbb{E}(E(\f,\g))
=\mathbb{E}(\f^t Q\,\g)
=\mathbb{E}(\f)^t Q\,\mathbb{E}(\g)
=(\boldsymbol{\mu}_f)^t Q\,(\boldsymbol{\mu}_g)
\,,
\end{equation*}
where $\f$ and $\g$ are assumed to be independent.
Furthermore,
\begin{align}
\mathbb{E}((E(\f,\g))^2)
&=\mathbb{E}((\f^tQ\,\g)^2)\nonumber
=\mathbb{E}((\f^tQ\,\g)(\f^tQ\,\g))\nonumber\\
&=\mathbb{E}((\g^tQ\,\f)(\f^tQ\,\g)) \quad \hbox{(as $Q$ is symmetric)}\nonumber\\
&=\mathbb{E}({\rm tr\!}\left[(\g^tQ\,\f)(\f^tQ\,\g)\right])\quad \hbox{(as the expression inside the square brackets is a scalar)}\nonumber\\
&=\mathbb{E}({\rm tr\!}\left[Q\,(\f\,\f^t)\,Q\,(\g\,\g^t)\right])\quad \hbox{(by the cyclic property of the trace)}\nonumber\\
&={\rm tr\!}\left[Q\,\mathbb{E}(\f\,\f^t)\,Q\,\mathbb{E}(\g\,\g^t)\right]
\quad \hbox{(as $\f$ and $\g$ are independent)}
\label{Esq}
\,.
\end{align}
Now, 
\begin{align*}
\mathbb{E}(\f\,\f^t)
&=\mathbb{E}((\f-\boldsymbol{\mu}_f)(\f-\boldsymbol{\mu}_f)^t +\boldsymbol{\mu}_f \f^t + \f(\boldsymbol{\mu}_f)^t-\boldsymbol{\mu}_f(\boldsymbol{\mu}_f)^t)\\
&=C_f+\boldsymbol{\mu}_f\,\mathbb{E}(\f)^t+\mathbb{E}(\f)\,(\boldsymbol{\mu}_f)^t-\boldsymbol{\mu}_f(\boldsymbol{\mu}_f)^t
=C_f+\boldsymbol{\mu}_f(\boldsymbol{\mu}_f)^t+\boldsymbol{\mu}_f(\boldsymbol{\mu}_f)^t-\boldsymbol{\mu}_f(\boldsymbol{\mu}_f)^t\\
&=C_f+\boldsymbol{\mu}_f(\boldsymbol{\mu}_f)^t
\end{align*}
and similarly,
\begin{equation*}
\mathbb{E}(\g\,\g^t)
=C_g+\boldsymbol{\mu}_g(\boldsymbol{\mu}_g)^t
\,.
\end{equation*}
Substituting these results for the means in  expression~(\ref{Esq}), we obtain the required formula.
The formula for the variance follows directly from the fact that
\begin{equation*}
{\rm var\!}\left[E(\f,\g)\right]
=\mathbb{E}((E(\f,\g))^2)-(\mathbb{E}(E(\f,\g)))^2
\,,
\end{equation*}
which completes the proof.\,$\square$\\

Let us consider specific cases of Lemma~\ref{Estats}.
\subsection{Deterministic $\f$}
Suppose that $\f$ is {\it fixed}, which means that $C_f=0$ and $\boldsymbol{\mu}_f=\f$\,.
Also, suppose that we have $n$ equally spaced layers, so that $w_k=1/n$\,, $1\leqslant k\leqslant n$\,.
For $\g$\,, we take $g_k \sim N(\mu_k,\sigma)$\,, where $1\leqslant k\leqslant n$\,, which is independent of $\f$\,, with $\boldsymbol{\mu}_g=[\mu_1\,,\mu_2\,,\cdots\,,\mu_n]^t$ and $C_g=\sigma^2 I_n\in\RR^{n\times n}$\,, where $\sigma$ is the standard deviation.

In this case, $E(\f,\g)=\f^tQ\,\g$\,, which is the sum of independent normal variables, is itself a normal random variable, whose mean and variance are given by Lemma~\ref{Estats}.
Specifically,
\begin{equation*}
\mathbb{E}(E(\f,\g))
=\f^tQ\,\boldsymbol{\mu}_g
=\overline{\f\,\boldsymbol{\mu}_g}-\overline{\f}\,\overline{\boldsymbol{\mu}_g}\,.
\end{equation*}
For the variance, and considering equally spaced weights, we have
\begin{equation*}
Q
=\frac{1}{n}I_n-\frac{1}{n^2}\mathbbm{1}_{n\times n}
\,,
\end{equation*}
where $\mathbbm{1}_{n\times n}\in\RR^{n\times n}$ denotes the matrix whose entries are all unity.
Then, since $C_f=0$\,, we have
\begin{align*}
{\rm var\!}\left[E(\f,\g)\right]
&={\rm tr\!}\left[Q(\f\,\f^t)\,Q\,(\sigma^2I_n)\right]
=\sigma^2\,{\rm tr\!}\left[Q\,(\f\,\f^t)\,Q\right]
=\sigma^2\,{\rm tr\!}\left[(Q\,\f)(Q\,\f)^t\right]\\
&=\sigma^2\,{\rm tr\!}\left[(Q\,\f)^t(Q\,\f)\right]\quad \hbox{(by the cylic property of the trace)}\\
&=\sigma^2 (Q\,\f)^t(Q\,\f)\quad \hbox{(as the expression in the square brackets above is just a scalar)}\,;
\end{align*}
but
\begin{equation*}
Q\,\f
=\left(\frac{1}{n}I_n-\frac{1}{n^2}\mathbbm{1}_{n\times n}\right)\f
=\frac{1}{n}(\f-\overline{\f}\,)\,,
\end{equation*}
so that
\begin{align*}
{\rm var\!}\left[E(\f,\g)\right]
&=\sigma^2\Bigl(\frac{1}{n^2}(\f-\overline{\f}\,)^t(\f-\overline{\f}\,)\Bigr)
=\frac{\sigma^2}{n^2}\sum_{k=1}^n (f_k-\overline{f}\,)^2
=\frac{\sigma^2}{n}\,\overline{(\f-\overline{\f}\,)^2}\,.
\end{align*}
In other words,
\begin{equation*}
{\rm std\!}\left[(E(\f,\g)\right]
=\frac{\sigma}{\sqrt{n}}\Bigl(\,\overline{(\f-\overline{\f}\,)^2}\,\Bigr)^{1/2}
\end{equation*}
is proportional to $\sigma$\,, which is the standard deviation of $g_k$\,, and inversely proportional to $\sqrt{n}$\,, where $n$ is the number of layers.
Thus, ${\rm std\!}\left[(E(\f,\g)\right]$ decreases with the number of layers; in other words, the approximation improves with the number of layers.
Since, in this case, $E(\f,\g)$ is a true normal variable, we expect that---with  $95\%$ probability---it is within two standard deviations of its mean, and with $99\%$ it is within $2.56$ standard deviations.
\section{Illustrative numerical examples}
\subsection{Introductory comments}
\label{sub:NonCon}
Let us remain within a medium composed of $n$ equally spaced layers, and let the thickness of the medium be $L=100$\,.
We consider the slowly moving wave, $f(x)=1+0.1\sin(2\pi x/100)$, passing through the medium, and model this wave by the piecewise constant vector given by the average of $f(x)$ on each layer.
In other words, 
\begin{align*}
f_k
=(\boldsymbol{\mu}_f)_k
&=\frac{1}{x_k-x_{k-1}}\int\limits_{x_{k-1}}^{x_k}\!\!f(x)\,{\rm d}x
=\frac{1}{h}\int\limits_{(k-1)h}^{kh}\!\!\left(1+0.1\sin\left(\frac{2\pi\,x}{100}\right)\right){\rm d}x\\
&=1+(0.1)\frac{1}{h}\frac{100}{2\pi}\left\{\cos\left(\frac{2\pi\left(k-1\right)h}{100}\right)-\cos\left(\frac{2\pi\,k\,h}{100}\right)\right\}\\
&=1+(0.1)\frac{n}{2\pi} \left\{\cos\left(\dfrac{2\pi\,(k-1)}{n}\right)-\cos\left(\dfrac{2\pi\,k}{n}\right)\right\}\\
&=1+(0.1)\frac{n}{2\pi}\left(2\sin\left(\dfrac{\pi}{n}\right)\sin\left(\dfrac{(2k-1)\,\pi}{n}\right)\right)\\
&=1+(0.1)\left(\frac{n}{\pi}\sin\left(\dfrac{\pi}{n}\right)\right)\sin\left(\dfrac{(2k-1)
\,\pi}{n}\right)
\end{align*}
and $C_f=0$\,, as $\f$ is deterministic.
Furthermore, we note that
\begin{equation*}
\overline{\f}=\overline{f(x)}=\frac{1}{L}\int\limits_0^L f(x)\,{\rm d}x=1
\,,
\end{equation*}
for any value of~$n$\,.
\subsection{Best case}
\label{sub:BestCase}
For the absolute error, the best possible situation is $\mathbb{E}(E(\f,\g))=E(\f,\boldsymbol{\mu}_g)=0$\,.
This is the case for any $\f$\,, if $\m_g\in\RR^n$ is a vector whose components $(\boldsymbol{\mu}_g)_k=\mu$\,, which is a constant; in such a case $\overline{\f\,\g}=\overline{\f}\,{\mu}=\overline{\f}\,\overline{\g}$\,.
Also, $E(\f,\m_g)=0$ if $\m_g$ that alternates between any two values, as can be verified by a calculation.

Let us suppose that the means of $g_k$\,, namely, $(\m_g)_k=\mu$\,, where $1\leqslant k\leqslant n$\,, are all the same.
Then, $\mathbb{E}(E(\f,\g))=0$\,, which means that, in this case, the expected difference between the mean of the product and the product of the means is zero.

Moreover, the proportionality constant in the variance of $E(\f,\g)$ becomes
\begin{align*}
\overline {(\f-\overline{\f}\,)^2}
&=\frac{1}{n}\sum_{k=1}^n (f_k-\overline{\f}\,)^2
=\frac{1}{n}\sum_{k=1}^n (f_k-1)^2\\
&=\frac{1}{n}(0.1)^2\left(\frac{n}{\pi}\sin\left(\dfrac{\pi}{n}\right)\right)^{\!2}\sum_{k=1}^n\left(\sin\left(\dfrac{(2k-1)\,\pi}{n}\right)\right)^{\!2}\\
&=\frac{1}{n}(0.1)^2\left(\frac{n}{\pi}\sin\left(\dfrac{\pi}{n}\right)\right)^{\!2}\sum_{k=1}^n\dfrac{1-\cos\left(2\,\dfrac{(2k-1)\,\pi}{n}\right)}{2}\\
&=\frac{(0.1)^2}{2}\left(\frac{n}{\pi}\sin\left(\frac{\pi}{n}\right)\right)^{\!2}\,,
\end{align*}
since
\begin{align*}
\sum_{k=1}^n\cos\left(2\dfrac{(2k-1)\,\pi}{n}\right)
&=\Re\left\{\sum_{k=1}^n \exp\left(\dfrac{2\,(2k-1)\,\pi\,i}{n}\right)
\right\}\\
&=\Re\left\{\exp\left(\dfrac{2\pi\,i}{n}\right)\sum_{k=1}^n\left(\exp\left(\dfrac{4\pi\,i}{n}\right)\right)^{\!k-1}\right\}\\
&=\Re\left\{\exp\left(\frac{2\pi\,i}{n}\right)\frac{\left(\exp\left(\dfrac{4\pi\,i}{n}\right)\right)^{\!n}-1}{\left(\exp\left(\dfrac{4\pi\,i}{n}\right)\right)-1}\right\}\\
&=0\,,
\end{align*}
for $n\geqslant 3$\,; herein, $\Re\{\,\}$ denotes the real part of a complex number.
Note that the factor 
\begin{equation*}
\left(\frac{n}{\pi}\sin\left(\frac{\pi}{n}\right)\right)^{\!2} \leqslant 1\,,
\qquad{\rm since}\qquad
\left|\frac{\sin(x)}{x}\right|\leqslant 1,\,\,\forall x\in\RR
\,.
\end{equation*}
Indeed,
\begin{equation*}
\left(\frac{\raisebox{6pt}{$n\sin\left(\dfrac{\pi}{n}\right)$}}{\pi}\right)^{\!\!2}
=1+O(n^{-2})
\,,
\end{equation*}
hence, it may be safely replaced by unity.

Thus,
\begin{equation*}
{\rm std\!}\left[E(\f,\g)\right]\le\sqrt{0.005}\frac{\sigma}{\sqrt{n}}=0.0707 \frac{\sigma}{\sqrt{n}}
\,,
\end{equation*}
and $95\%$ of the time $E(\f,\g)$ is in the interval between
\begin{equation*}
\pm 2\,(0.0707)\frac{\sigma}{\sqrt{n}}
=\pm 0.1414\frac{\sigma}{\sqrt{n}}
\,.
\end{equation*}

\begin{figure}
\begin{center}
\includegraphics[scale=0.5]{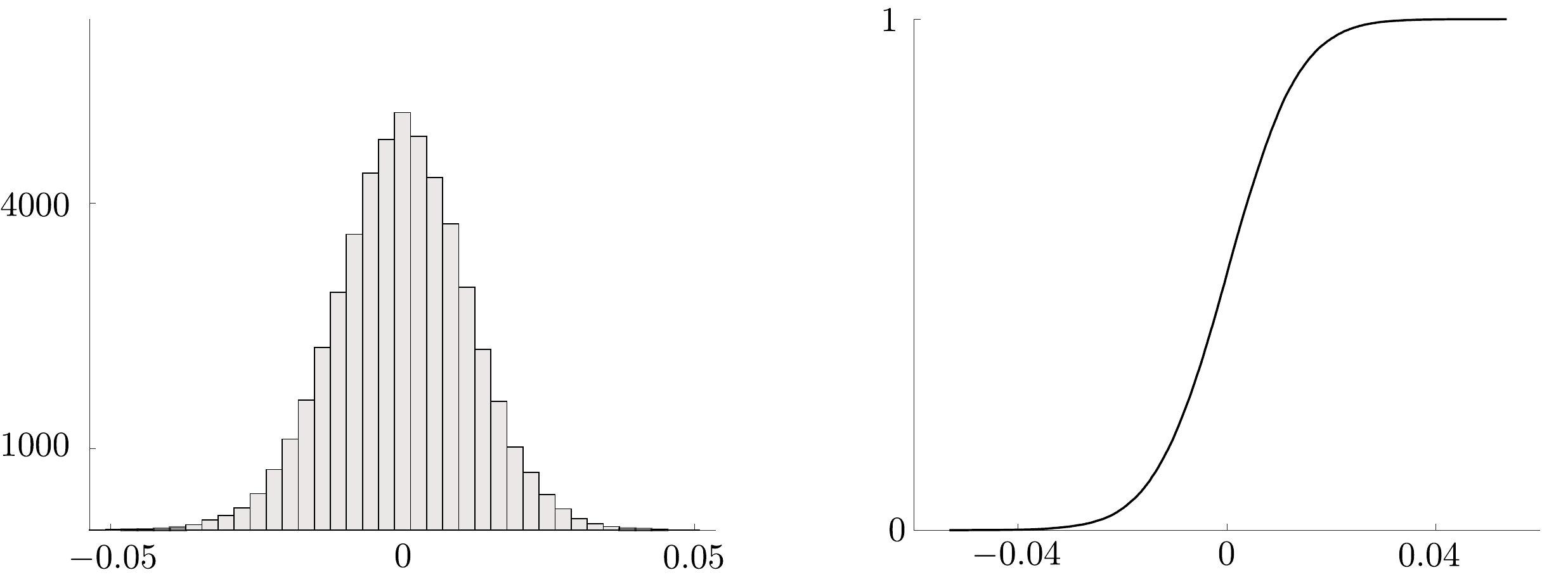}
\end{center}
\caption{Relative errors of $\overline{\f\,\g}\approx \overline \f\,\overline \g$ for the case discussed in Section~\ref{sub:BestCase}}
\label{fig:FigRelErrs1}
\end{figure}

The {\it relative} errors, defined by
\begin{equation}
\label{eq:Rfg}
R(\f,\g)=\frac{E(\f,\g)}{\overline{\f\,\g}}
\,,
\end{equation}
are another issue, since they are a ratio of two random variables.
Information about them can be obtained by generating a number of simulations.
In Figure~\ref{fig:FigRelErrs1}, we show the results for fifty thousand simulations with $\mu=2$\,, $\sigma=1$ and $n=10$\,.
In this and the other figures, both the left and right plots contain essentially the same information.
The left plot is a histogram of the number of occurrences corresponding to a given value, and the right plot is their cumulative sum normalized to unity.

Also, for these simulations, we obtain the following results.
\renewcommand{\labelitemi}{$\bullet$}
\begin{itemize}
\item[$\circ$] $95\%$ of $|E(\f,\g)|$ are less than $0.0432$
\item[$\circ$] $99\%$ of $|E(\f,\g)|$ are less than $0.0564$
\item[$\circ$] the maximum of $|E(\f,\g)|$ is $0.0875$
\item[$\circ$] $95\%$ of $|R(\f,\g)|$ are less than $2.2611\%$
\item[$\circ$] $99\%$ of $|R(\f,\g)|$ are less than $3.1173\%$
\item[$\circ$] the maximum of $|R(\f,\g)|$ is $5.3633\%$
\item[$\circ$] the theoretical mean of $E(\f,\g)$ is $0.0000$ 
\item[$\circ$] the sample mean of $E(\f,\g)$ is $0.0000$ 
\item[$\circ$] the theoretical standard deviation of $E(\f,\g)$ is $2.1995\times10^{-2}$
\item[$\circ$] the sample standard deviation of $E(\f,\g)$ is $2.1950\times10^{-2}$
\end{itemize}

\medskip \noindent {\bf Remark.}
Large relative errors are typically caused by the division of a small value of $\overline{\f\,\g}$\,.
Note that, in general, we may write
\begin{equation*}
\overline{\f\,\g}
=\f^t\widetilde{Q}\,\g\,
\qquad\hbox{where}\qquad
\widetilde{Q}
:=\frac{1}{n}I_n\in\RR^{n\times n}
\,,
\end{equation*}
and, hence, if $\f$ is fixed and $g_k\sim N((\m_g)_k,\sigma)$\,, we may invoke Lemma~\ref{Estats} to compute
\begin{equation*}
\mathbb{E}(\,\overline{\f\,\g}\,)
=\overline{\f\,\m_g}\,
\qquad\hbox{and}\qquad
{\rm std\!}\left[\,\overline{\f\,\g}\,\right]
=\frac{\sigma}{\sqrt{n}}(\,\overline{\f^2}\,)
\,.
\end{equation*}
We should expect the vast majority of values of $\overline{\f\,\g}$ to lie in the interval between 
\begin{equation*}
\mathbb{E}(\,\overline{\f\,\g}\,)\pm 2\,{\rm std\!}\left[\,\overline{\f\,\g}\,\right]
=\overline{\f\,\m_g}\pm 2\frac{\sigma}{\sqrt{n}}(\,\overline{\f^2}\,)
\,.
\end{equation*}
If this interval includes zero, then there are likely to be many instances for which  $\overline{\f\,\g}$ is small, and hence, the resulting relative error is large.
\medskip

However, in the case under consideration, $\mathbb{E}(\,\overline{\f\,\g}\,)=2\,\overline{\f}=2$\,, while ${\rm std\!}\left[\,\overline{\f\,\g}\,\right]=\overline{\f^2}/\sqrt{n}=0.3178$\,.
Hence, it is essentially impossible for a sample $\overline{\f\,\g}$ to be near zero and be the cause of a large relative error.
\subsection{Worst case}
\label{sub:WorstCase}
Let us now consider an almost worst case, for which the expected value of $E(\f,\g)$ is not zero.
Specifically, we consider  $\boldsymbol{\mu}_g=c\,(\f-\overline{\f})$\,, so that the upper bound given in Lemma~\ref{Ebnd} is attained.
In such a case,
\begin{equation*}
\overline{\m}_g
=c\,(\,\overline{\f-\overline{\f}}\,)
=c\,(\,\overline{\f}-\overline{\f}\,)
=0
\,,
\end{equation*}
and hence,
\begin{equation*}
E(\f,\m_g)
=\overline{\f\,\m}_g
\,;
\end{equation*}
more importantly,
\begin{equation*}
R(\f,\m_g)=\frac{E(\f,\m_g)}{\overline{\f\,\m}_g}=1
\,,
\end{equation*}
which means that the relative error is $100\%$\,.

\begin{figure}
\begin{center}
\includegraphics[scale=0.5]{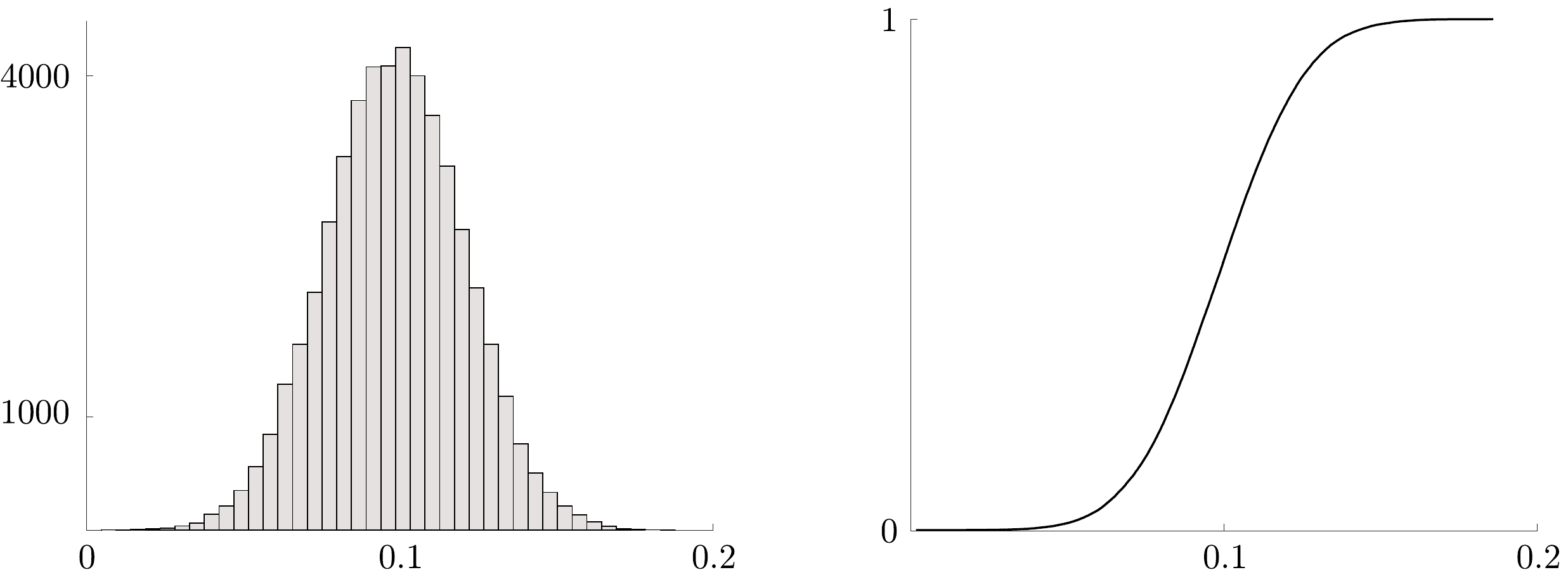}
\end{center}
\caption{Absolute errors of $\overline{\f\,\g}\approx \overline \f\,\overline \g$ for the case discussed in Section~\ref{sub:WorstCase}}
\label{fig:FigAbsErrs2}
\end{figure}

\begin{figure}
\begin{center}
\includegraphics[scale=0.5]{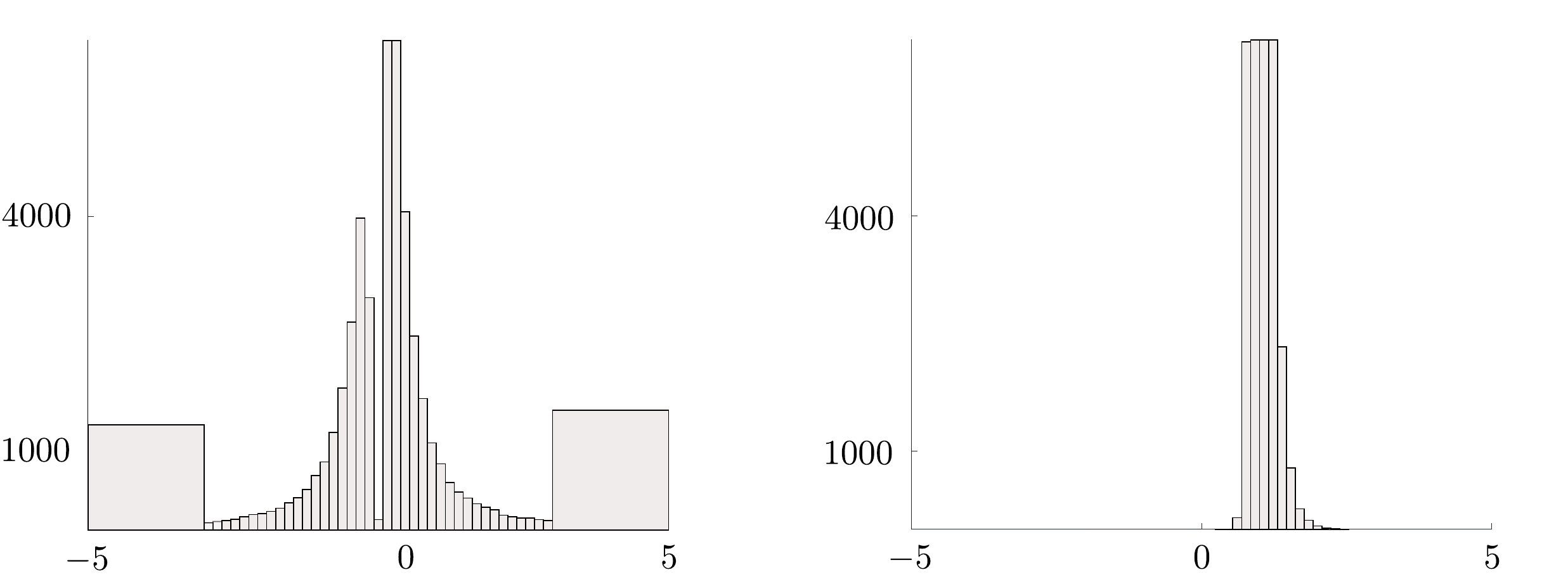}
\end{center}
\caption{Relative errors of $\overline{\f\,\g}\approx \overline \f\,\overline \g$ for the case discussed in Section~\ref{sub:WorstCase}; left plot: $\sigma=1$\,; right plot: $\sigma=0.05$}
\label{fig:FigRelErrs23}
\end{figure}

Specifically, we take $g_k\sim N(\mu_k,\sigma)$\,, independent, with $\mu_k=2\sin((2k-1)\pi/n)$ and we set $\sigma=1$\,.
Since $E(\f,\g)$ is still a normal random variable, it behaves as illustrated in Figure~\ref{fig:FigAbsErrs2}.
Indeed, the standard deviation of $E(\f,\g)$ is the same as for case discussed in Section~\ref{sub:BestCase}, since it does not depend on $\boldsymbol{\mu}_g$\,; however,
\begin{equation*}
\mathbb{E}(E(\f,\g))=E(\f,\boldsymbol{\mu}_g)=0.0984
\,.
\end{equation*}
The relevant statistics for the absolute errors are as follows.
\begin{itemize}
\item[$\circ$] $95\%$ of $|E(\f,\g)|$ are less than $0.1343$
\item[$\circ$] $99\%$ of $|E(\f,\g)|$ are less than $0.1499$
\item[$\circ$] the maximum of $|E(\f,\g)|$ is $0.1860$
\item[$\circ$] the theoretical mean of $E(\f,\g)$ is $0.0984$
\item[$\circ$] the sample mean of $E(\f,\g)$ is $0.0984$
\item[$\circ$] the theoretical standard deviation of  $E(\f,\g)$ is $2.1995\times 10^{-2}$ 
\item[$\circ$] the sample standard deviation of the  $E(\f,\g)$ is $2.2020\times 10^{-2}$
\end{itemize}
However, the relative error,~$R(\f,\g)$\,, is almost catastrophically worse.
Examining Figure~\ref{fig:FigRelErrs23}, we see the frequencies of the relative errors for fifty thousand simulations.
Notice that, in the left plot, there are many cases for which the relative error exceeds $100\%$\,.
Indeed, this is true for  $23.24\%$ of these simulations.
Even $11.95\%$ of them are over $200\%$\,, and only $1.17\%$ of the relative errors are below $10\%$\,. 

Such a magnitude of relative errors is easy to explain.
Since $R(\f,\m_g)=1$\,, we should expect relative errors to be typically around $100\%$\,.
Also, if $\overline{\f\,\g}$ is small---which is possible for small $n$ and  $\sigma$\,, since the standard deviation of $g_k$ is large relative to $\mathbb{E}(\f\,\g)$---the division by the small number amplifies the relative error, as is the case herein. 

To illustrate this effect, we repeat the same experiment, except with $\sigma=0.05$\,, as opposed to $\sigma=1$\,.
The result is shown in the right plot of Figure~\ref{fig:FigRelErrs23}.
Comparing the left and right plots, we see that---for  $\sigma=0.05$---the relative errors are much more concentrated around the expected value of $100\%$\,, since it is much less likely that $\overline{\f\,\g}$ would be small.
\subsection{Intermediate case}
\label{sub:InterCase}
\begin{figure}
\begin{center}
\includegraphics[scale=0.5]{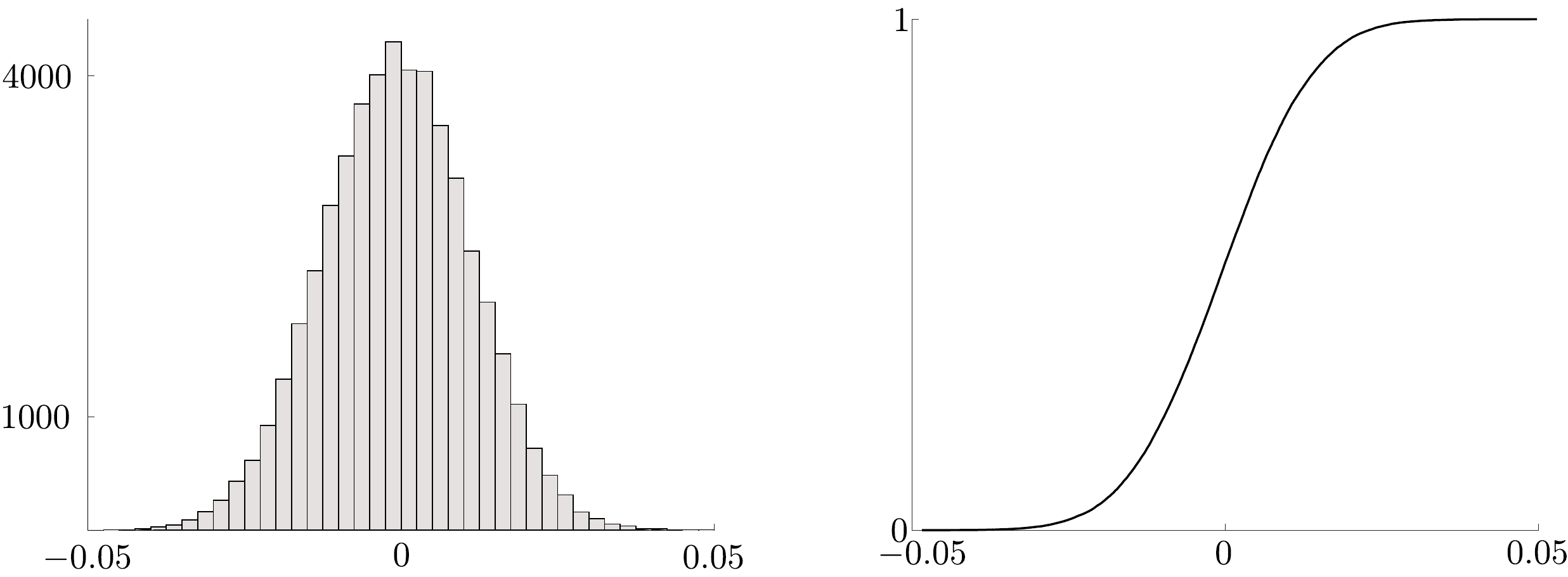}
\end{center}
\caption{Absolute errors of $\overline{\f\,\g}\approx \overline \f\,\overline \g$ for the case discussed in Section~\ref{sub:InterCase}, with $\sigma=0.75$}
\label{fig:FigAbsErrs6}
\end{figure}

\begin{figure}
\begin{center}
\includegraphics[scale=0.5]{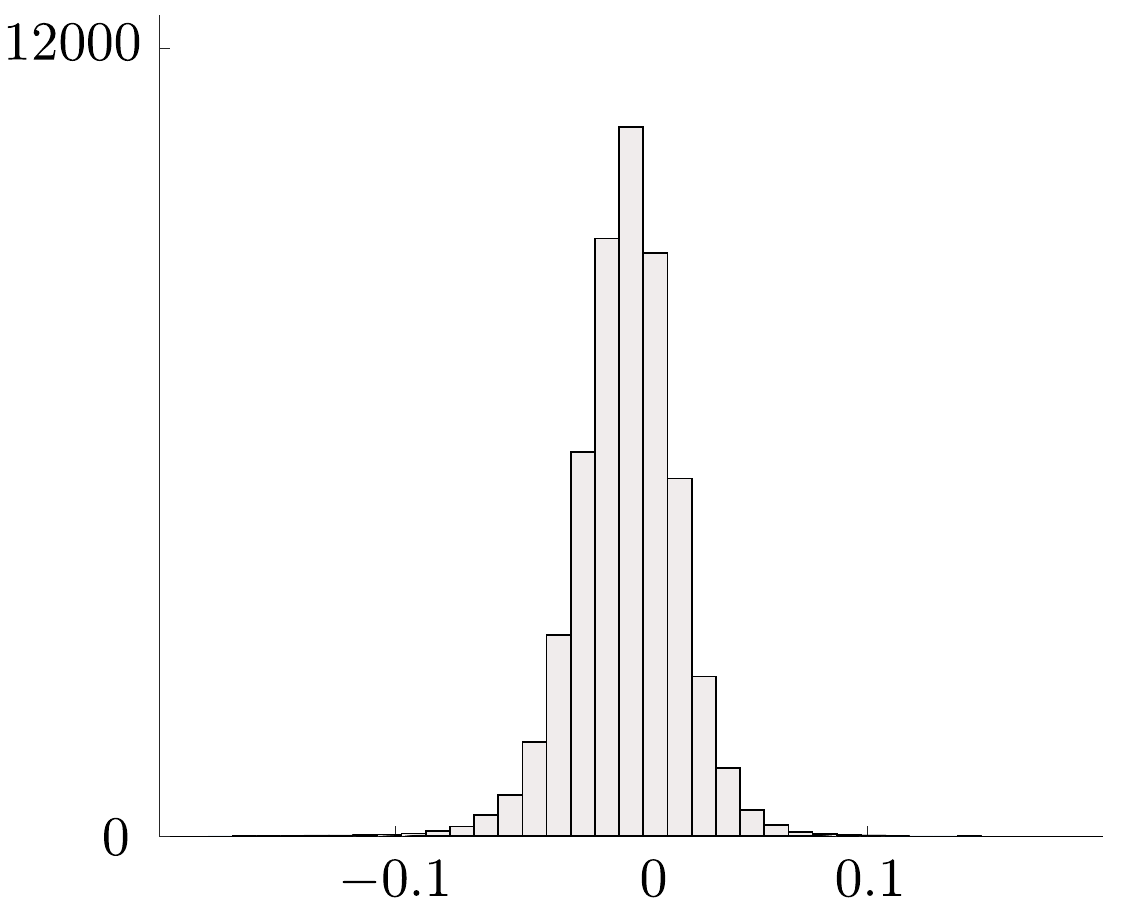}
\end{center}
\caption{Relative errors of $\overline{\f\,\g}\approx \overline \f\,\overline \g$ for the case discussed in Section~\ref{sub:InterCase}, with $\sigma=0.75$}
\label{fig:FigRelErrs6}
\end{figure}

Having examined the best and worst cases, let us consider an intermediate one.
To do so, we set $g$ to represent typical values to which the Backus~\cite{Backus1962} average is applied (e.g., Danek and Slawinski~\cite{DanekSlawinski2016}).
We use the same $\f$ as for the cases discussed in Sections~\ref{sub:BestCase} and \ref{sub:WorstCase}.
For $g$\,, we consider twenty isotropic layers of even thickness, whose elasticity parameters are either $c_{1111}=12.15$ and $c_{2323}=3.24$ or $c_{1111}=6.25$ and $c_{2323}=0.64$\,.
For each layer, the value of $g$ is given by $(c_{1111}-2c_{2323})/c_{1111}$\,, which is the term in parentheses of expression~(\ref{eq:WS2016}).
The sequence of layers is random; the same pair of values can be repeated, which is tantamount to doubling the thickness of a layer.
The step function,~$g$\,, and, hence, $\m_g$\,, alternate between $0.4667$ and $0.7952$\,.
Herein, we consider
\begin{align*}
\m_g&=[0.4667, 0.7952, 0.7952, 0.4667, 0.7952, 0.4667, 0.4667, 0.4667, 0.7952, 0.7952,\\ 
    &\qquad 0.7952, 0.4667, 0.4667, 0.4667, 0.7952, 0.4667, 0.7952, 0.7952, 0.7952, 0.4667]\in \RR^{20}.
\end{align*}
As in the cases examined in Sections~\ref{sub:BestCase} and \ref{sub:WorstCase}, we take
\begin{equation*}
\g_k\sim N((\m_g)_k,\sigma),\qquad 1\leqslant k\leqslant n
\,,
\end{equation*}
but with $n=20$ and $\sigma=0.75$\,.
The results of fifty thousand simulations are shown in Figures~\ref{fig:FigAbsErrs6} and \ref{fig:FigRelErrs6}.
The relevant statistics are as follows.
\begin{itemize}
\item[$\circ$] $95\%$ of the $|E(\f,\g)|$ are less than $0.0232$
\item[$\circ$] $99\%$ of the  $|E(\f,\g)|$ are less than $0.0305$
\item[$\circ$] the maximum of the $|E(\f,\g)|$ is $0.0498$
\item[$\circ$] $95\%$ of  $|R(\f,\g)|$ are less than $4.2930\%$
\item[$\circ$] $99\%$ of $|R(\f,\g)|$ are less than $6.6116\%$
\item[$\circ$] the maximum of $|R(\f,\g)|$ is $108.3342\%$
\item[$\circ$] the theoretical mean of $E(\f,\g)$ is $-0.0007$ 
\item[$\circ$] the sample mean of $E(\f,\g)$ is $-0.0008$ 
\item[$\circ$] the theoretical standard deviation of $E(\f,\g)$ is $1.1810\times10^{-2}$ 
\item[$\circ$] the sample standard deviation of $E(\f,\g)$ is $1.1793\times10^{-2}$ 
\item[$\circ$] the theoretical mean of $\overline{\f\,\g}$ is $0. 6302$
\item[$\circ$] the theoretical standard deviation of $\overline{\f\,\g}$ is $0.1685$
\end{itemize}
Notice that the expected value of  $\overline{\f\,\g}$ is $0.6302$\,, while its standard deviation is $0.1685$\,, which means that a small value for $\overline{\f\,\g}$ is {\it possible} but not {\it likely}.
We see this illustrated by the distribution of the relative errors,~$R(\f,\g)$\,, for which $99\%$ of its values are less than $6.6116\%$\,, in absolute value, while its maximum absolute value is as large as $108.3342\%$\,.
\subsection{Effect of measurement errors}
\label{sec:ErrorEffects}
We may also use our formulation to study the effect of small random errors in the values of the $f_k$ and the $g_k$\,.
Since, in general, there is no analytic expression for error propagation, we use numerical methods to gain an insight into the effect of measurement errors.
To this end, we introduce random normal errors of 10\% to $\f$ and $\g$\,.
Specifically, in accordance with Section~\ref{sub:NonCon}, we let the mean, $\boldsymbol{\mu}_f\in\RR^n$\,, be
\begin{equation*}
(\boldsymbol{\mu}_f)_k
:= \frac{1}{x_{k+1}-x_k}\int\limits_{x_k}^{x_{k+1}}\!\!f(x)\,{\rm d}x
\,,
\end{equation*}
but we consider
$f_k=(\boldsymbol{\mu}_f)_k+(\boldsymbol{\mu}_f)_k\,\sigma\,z_k$\,, where $z_k\sim N(0,1)$\,, with $1\leqslant k\leqslant n$\,,
are independent, and $\sigma=0.1$\,.
In other words,
\begin{equation*}
f_k\sim N\left((\boldsymbol{\mu}_f)_k,(\sigma\, (\boldsymbol{\mu}_f)_k)^2\right)
\,,
\end{equation*}
and, hence, the correlation matrix is
\begin{equation*}
(C_f)_{ij}=\mathbb{E}\left(\left(f_i-(\boldsymbol{\mu}_f)_i\right)\left(f_j-(\boldsymbol{\mu}_f)_j\right)\right)
=
\begin{cases}
(\sigma\, (\boldsymbol{\mu}_f)_j)^2
&\hbox{if}\,\,i=j\cr0
&\hbox{otherwise}
\end{cases}\,.
\end{equation*}
In other words,
\begin{equation*}
C_f
=\sigma^2\,{\rm diag\!}\left[(\boldsymbol{\mu}_f)^2\right]\in\RR^{n\times n}
\,,
\end{equation*}
which is a diagonal matrix whose entries are $(\boldsymbol{\mu}_f)_j^2$\,.
Similarly we take 
\begin{equation*}
g_k
\sim N\left(\left(\boldsymbol{\mu}_g)_k,(\sigma\, (\boldsymbol{\mu}_g)_k\right)^2\right)
\,,
\end{equation*}
for which 
\begin{equation*}
C_g
=\sigma^2\,{\rm diag\!}\left[(\boldsymbol{\mu}_g)^2\right]\in\RR^{n\times n}
\,,
\end{equation*}
which is a diagonal matrix whose entries are $(\boldsymbol{\mu}_g)_j^2$\,.

\begin{figure}
\begin{center}
\includegraphics[scale=0.5]{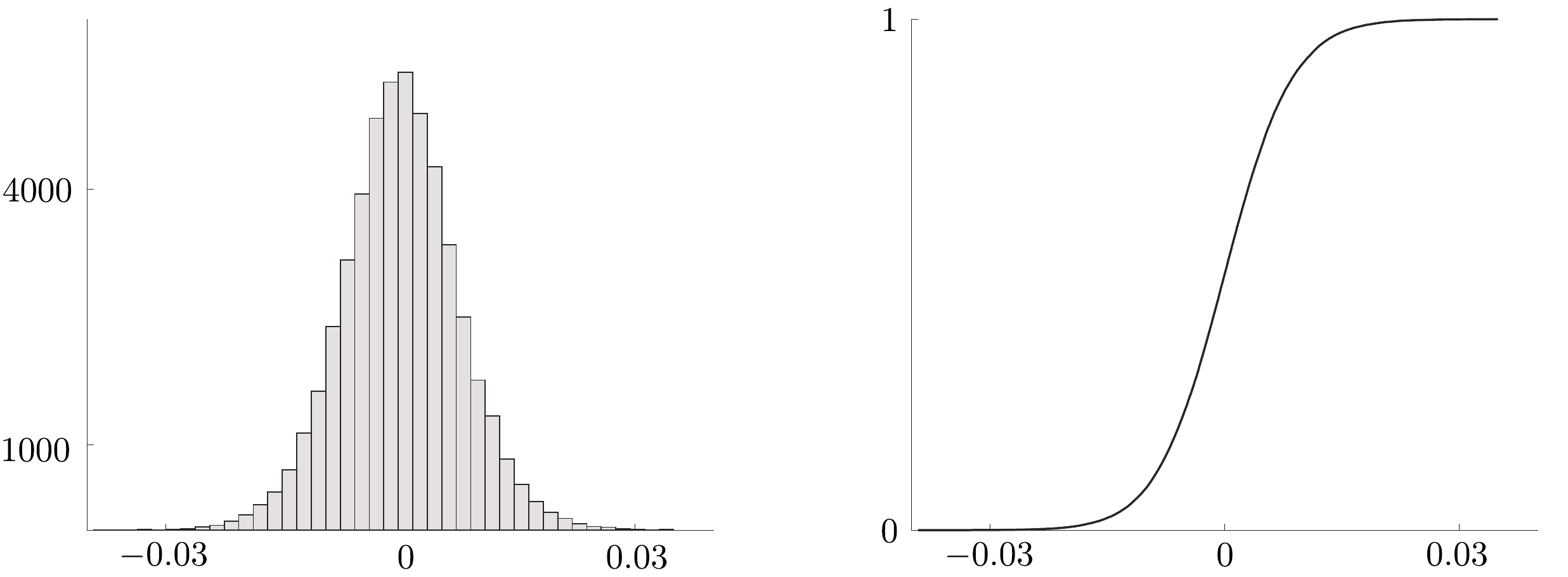}
\end{center}
\caption{Absolute errors of $\overline{\f\,\g}\approx \overline \f\,\overline \g$ for $10\%$ errors in $\f$ and $\g$  }
\label{fig:FigAbsErrs4}
\end{figure}

\begin{figure}
\begin{center}
\includegraphics[scale=0.5]{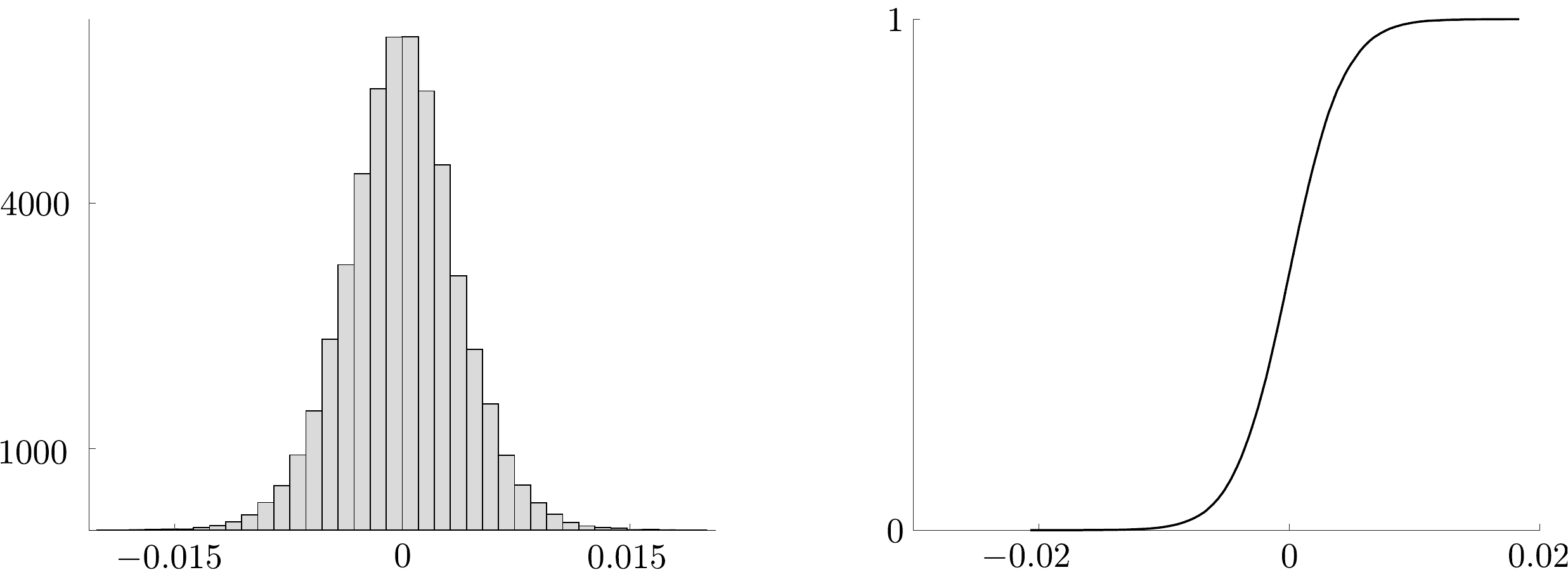}
\end{center}
\caption{Relative errors of $\overline{\f\,\g}\approx \overline \f\,\overline \g$ for $10\%$ errors in $\f$ and $\g$ for $n=10$}
\label{fig:FigRelErrs4}
\end{figure}

According to Lemma~\ref{Estats}, $\mathbb{E}(E(\f,\g))=E(\boldsymbol{\mu}_f,\boldsymbol{\mu}_g)$\,; also, its variance is given therein.
From this, it follows that
${\rm std\!}\left[E(\f,\g)\right]$ is again  proportional to $\sigma/\sqrt{n}$\,.
We note that, in this case, $E(\f,\g)$ is {\it not} a normal random variable, being the sum of {\it products} of normal variables.
In Figures~\ref{fig:FigAbsErrs4} and \ref{fig:FigRelErrs4} we show the results for the case of ten layers and $(\boldsymbol{\mu}_g)_k=2$\,, where $1\leqslant k \leqslant 10$\,.
We notice that the errors are very reasonably behaved.
The relevant statistics are as follows.
\begin{itemize}
\item[$\circ$] $95\%$ of $|E(\f,\g)|$ are less than $0.0149$
\item[$\circ$] $99\%$ of $|E(\f,\g)|$ are less than $0.0208$
\item[$\circ$] the maximum of $|E(\f,\g)|$ is $0.0392$
\item[$\circ$] $95\%$ of $|R(\f,\g)|$ are less than $0.7488\%$
\item[$\circ$] $99\%$ of $|R(\f,\g)|$ are less than $1.0357\%$
\item[$\circ$] the maximum of $|R(\f,\g)|$ is $2.0653\%$
\item[$\circ$] the theoretical mean of $E(\f,\g)$ is $0$ 
\item[$\circ$] the sample mean of $E(\f,\g)$ is $0$ 
\item[$\circ$] the theoretical standard deviation of $E(\f,\g)$ is $7.4515\times10^{-3}$
\item[$\circ$] the sample standard deviation of $E(\f,\g)$ is $7.4627\times10^{-3}$
\end{itemize}

\begin{figure}
\begin{center}
\includegraphics[scale=0.5]{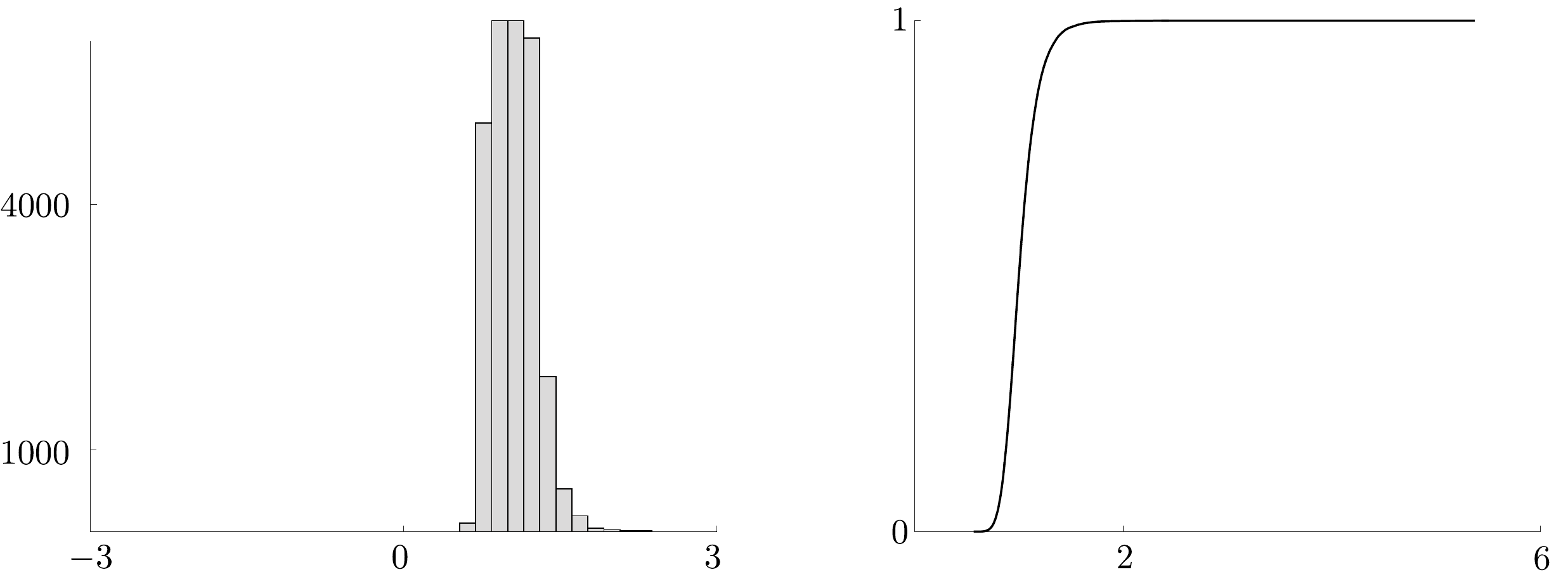}
\end{center}
\caption{Relative errors of $\overline{\f\,\g}\approx \overline \f\,\overline \g$ for $10\%$ errors in $\f$ and $\g$ for $n=100$  }
\label{fig:FigRelErrs5}
\end{figure}

If the same procedure is applied to the case discussed in Section~\ref{sub:WorstCase}, the absolute errors $E(\f,\g)$ behave in a similar manner, but the relative errors are large, as expected.
Since, in this case, $\sigma$ is indicative of the level of numerical ``noise'' in the data, it is not likely or reasonable that it be reduced.
However, we note that---since the standard deviation is inversely proportional to $\sqrt{n}$\,---the larger the value of~$n$\,, the smaller the relative errors.
Examining Figure~\ref{fig:FigRelErrs5}, we see that the relative errors are clustered around the expected value of $100\%$\,.
The relevant statistics for the relative error are as follows.
\begin{itemize}
\item[$\circ$] $95\%$ of $|R(\f,\g)|$ are less than $131.88\%$
\item[$\circ$] $99\%$ of $|R(\f,\g)|$ are less than $154.07\%$
\item[$\circ$] the maximum of $|R(\f,\g)|$ is $536.8065\%$
\end{itemize}
\subsection{$\overline{g}\approx0$ case}\label{sub:Hooke}
If $\overline{g}=0$\,, then, according to expression~(\ref{eq:E(f,g)}), $E(\f,\g)=\overline{\f\,\g}$\, and, hence, in accordance with expression~(\ref{eq:Rfg}), $R(\f,\g)=100\%$\,.
The relative errors are then amplified catastrophically if $\overline{\f\,\g}\approx0$\,.
Let us briefly discuss the specifics of such a situation.
In a manner similar to Section~\ref{sub:NonCon}, we let
\begin{equation*}
f(x)=1+a\sin\left(\frac{2\pi\,x}{L}\right)
\,,
\end{equation*}
which oscillates around its mean value of unity with the amplitude of~$a$ and the wavelength of~$L$\,.
If $\overline{\f\,\g}=0$\,, then
\begin{align*}
0
&=\sum_{k=1}^n w_k\,f_k\,g_k
=\sum_{k=1}^n w_k\,g_k + \sum_{k=1}^nw_k(f_k-1)\,g_k
=\overline{\g}+\sum_{k=1}^nw_k(f_k-1)\,g_k\\
&\approx \overline{\g}+\frac{1}{L}\int\limits_0^L (f(x)-1)\,g(x)\,{\rm d}x
= \overline{\g}+\frac{a}{L}\int\limits_0^L\sin\left(\frac{2\pi\,x}{L}\right)g(x)\,{\rm d}x\,.\\
\end{align*}
Consequently,
\begin{equation*}
\overline{\g}
\approx -\frac{a}{L}\int\limits_0^L\sin\left(\frac{2\pi\,x}{L}\right)g(x)\,{\rm d}x
=-\frac{a}{2\pi}\int\limits_0^{2\pi}\sin(x)\,g\!\left(\frac{L\,x}{2\pi}\right){\rm d}x
\,.
\end{equation*}
It follows that, in general,
\begin{equation*}
|\overline{\g}|
\lesssim\frac{a}{2\pi}\int\limits_0^{2\pi}\,\left|g\!\left(\frac{L\,x}{2\pi}\right)\right|\,{\rm d}x\leqslant a\max_{0\leqslant x\leqslant L}|g(x)|
\,,
\end{equation*}
is bounded proportionally to the amplitude,~$a$\,, and hence must be small; herein, $\lesssim$ stands for approximately~$\leqslant$\,.
Note that if $g(x)$ is a step function, we have $\leqslant$\,; otherwise, we have $\lesssim$\,, in general.

Also note that
\begin{equation*}
\frac{1}{\pi}\int\limits_0^{2\pi}\sin(x)\,g\!\left(\frac{L\,x}{2\pi}\right){\rm d}x
\end{equation*}
is the Fourier coefficient of $\sin(x)$\,, with unit frequency, for $g(Lx/(2\pi))$\,.
If $g(x)$ is rapidly varying or has a small component of unit frequency, then this coefficient is small, thus forcing
$|\overline{\g}|$ to be the product of two small numbers, which is very small.
Thus, we expect the problematic case of large relative error to occur only if $\overline{\g}$ is near zero.

Let us illustrate the case of $\overline g=0$ in the Backus~\cite{Backus1962} average within the context of layers composed of isotropic Hookean solids.
In such a case, expression~(\ref{eq:Hooke}) is reduced to
\begin{equation*}
\sigma_{ij}=\left(c_{1111}-2c_{2323}\right)\delta_{ij}\sum_{k=1}^{3}\varepsilon_{kk}+2c_{2323}\,\varepsilon_{ij}\,,\qquad i,j=1,2,3\,,
\end{equation*}
where $\delta_{ij}$ is the Kronecker delta.
Thus, we need to consider only two elasticity parameters: $c_{1111}$ and $c_{2323}$\,. 
Following details of the derivation presented by Slawinski~\cite[Section~4.2.2.2]{SlawinskiWS2016}, we consider the expression given by
\begin{equation}
\label{eq:WS2016}
\overline{\left(\frac{c_{1111}-2c_{2323}}{c_{1111}}\right)\frac{\partial}{\partial x_1}u_1}
\,,
\end{equation}
where $u_1$ is a component of the displacement vector, whose partial derivative with respect to $x_1$ is a component of the strain tensor,~$\varepsilon_{11}$\,.
The same form of expression also appears with $\partial u_2/\partial x_2=:\varepsilon_{22}$\,.
These are the two cases that can result in $\overline g=0$\,.
Other forms appearing in the derivation, such as $\overline{(1/c_{1111})\,\sigma_{33}}$ cannot lead to that result.

Following the Backus~\cite{Backus1962} approach, we approximate the average of a product by the product of their averages.
Assuming that one of the factors varies slowly, we
approximate expression~(\ref{eq:WS2016}) by
\begin{equation*}
\overline{\!\left(\frac{c_{1111}-2c_{2323}}{c_{1111}}\right)}\,\overline{\varepsilon_{11}}
\,,
\end{equation*}
where $\varepsilon_{11}:=\partial u_1/\partial x_1$\,.
This strain-tensor component is assumed to be nearly constant; within this paper, it corresponds to $f$\,.
The term composed of elasticity parameters, on the other hand, can be a rapidly varying function, which corresponds to $g$\,.

Stability conditions of Hookean solids, which are expressed as the positive definiteness of the elasticity tensor, require that both $c_{1111}$ and $c_{2323}$ be positive.
Thus, if $c_{1111}>2\,c_{2323}$\,, for all layers, $g$ is positive for all~$x$\,.
If, in any layer, $\tfrac{4}{3}\,c_{2323}<c_{1111}<2\,c_{2323}$\,, $g$ is negative in that layer.
The lower limit is also required by the stability conditions.

The range of the elasticity parameters resulting in negative values of $g$ appears to be less common in modelling natural materials.
It corresponds to Hookean solids exhibiting high rigidity.
Expressed in terms of $\alpha$ and $\beta$\,, which are the $P$-wave and $S$-wave speeds, respectively, the negative values occur if and only if
\begin{equation*}
2\beta/\sqrt 3<\alpha<\sqrt 2\beta
\,.
\end{equation*}
The lower limit is the closest allowable case of the two speeds.
The upper limit is still below the case of the so-called Poisson's solid, whose $\alpha=\sqrt 3\,\beta$\,; for such a solid, the Poisson ratio is $1/4$\,, and the two Lam\'e parameters are equal to one another.

Poisson's solid is representative of common sedimentary rocks.
Thus, the change of sign for the term composed of elasticity parameters, although it might occur, appears to be limited to values that are not common for seismic measurements in sedimentary basins.
Therein, the values of the quickly varying function are expected to remain positive.
\section{Conclusions}
The formulation presented in this paper provides tools that allow for the examination of the errors in approximation~(\ref{eq:Lemma1}), namely, $\overline{f\,g}\approx \overline f\,\overline g$\,, which is crucial for Backus~\cite{Backus1962} averaging.
If one considers only the upper bound, given previously by Bos et al.~\cite{BosEtAl2017}, Backus~\cite{Backus1962} averaging might not appear as a viable approach.
Yet, as demonstrated in this paper, for cases representative of physical  scenarios modelled with such an averaging, the approximation is reasonable.

Only the case of $\overline{\g}\approx 0$\,, where $g$ is the quickly varying function that represents properties of Hookean layers, raises concerns with respect to large relative errors.
However, as discussed in Section~\ref{sub:Hooke}, for sedimentary layers---which is a common scenario for the application of the Backus average---  $\overline{\g}\approx 0$ is unlikely to occur, since it would require the value of the term in parentheses of expression~(\ref{eq:WS2016}) to exhibit both positive and negative values within the region considered by the averaging process.
While  positive values are common in the Earth's crust,  negative values appear in the Earth's inner core (Prescher et al.~\cite{Poisson}), where the Hookean model of  the core approaches the maximum allowable value of Poisson's ratio,~$1/2$\,, which corresponds to $\alpha=2\beta/\sqrt 3$\,.
Thus, since the positive and negative values are unlikely to occur together in the same region within the Earth, the problematic issue of approximation~(\ref{eq:Lemma1}) is not likely to appear in seismology.
It might, however, appear in other aspects of material sciences where Backus~\cite{Backus1962} averaging might be applied.

The case of $\overline{\g}\approx 0$ might also occur for anisotropic layers discussed by Bos et al.~\cite{BosEtAl2017}.
For such cases, there are more expressions analogous to the fractional term in expression~(\ref{eq:WS2016}), as exemplified for orthotropic layers by Slawinski~\cite[Exercise~4.6]{SlawinskiWS2016}.
However, the stability conditions for anisotropic solids form a set of complicated inequalities and tend to prevent changes of sign of these expressions that would lead to $\overline{\g}=0$\,.
Hence, approximation~(\ref{eq:Lemma1}) remains reasonable for anisotropic layers.
\section*{Acknowledgments}
We wish to acknowledge discussions with David Dalton, Andrey Melnikov and Michael Rochester, the graphic support of Elena Patarini as well as the insightful comments of Alexey Stovas and Yuriy Ivanov, who refereed this paper.
This research was performed in the context of The Geomechanics Project supported by Husky Energy.
Also, this research was partially supported by the Natural Sciences and Engineering Research Council of Canada, grant 238416-2013, and by the Polish National Science Center under contract No.\ DEC-2013/11/B/ST10/0472.
\bibliographystyle{amsplain}
\bibliography{2018BDtSS_arXiv.bib}
\setcounter{section}{0}
\setlength{\parskip}{0pt}
\renewcommand{\thesection}{Appendix~\Alph{section}}
\section{Backus-average product approximation
(Lemma~\ref{lem:Len})}
\label{app:Len}
To discuss the details of the upper bound of the Backus-average product approximation, let us consider the following.
\begin{align*}
\left|\overline{f\,g}-\overline{f}\,\overline{g}\right|
=\left|f(c)-\overline{f}\right|\overline{g}
&=\left|\,\int\limits_{-\infty}^{\infty}f(c)\,W(\zeta)\,{\rm d}\zeta - \int\limits_{-\infty}^{\infty}f(\zeta)\,W(\zeta)\,{\rm d}\zeta\,\right|\,\overline{g} \\
&=\left|\,\int\limits_{-\infty}^{\infty}\left(f(c)-f(\zeta)\right)W(\zeta)\,{\rm d}\zeta\,\right|\,\overline{g} \\
&\leqslant\left\{\int\limits_{-\infty}^{\infty}\left|f(c)-f(\zeta)\right|W(\zeta)\,{\rm d}\zeta\,\right\}\,\overline{g}\,,
\end{align*}
where, for a fixed $x$\,, $W(\zeta):=w(\zeta-x)$\,.
By the Mean Value Theorem for derivatives
\begin{equation*}
f(c)-f(\zeta)
= f'(a)(c - \zeta)
\end{equation*}
for some intermediate $a$ between $c$ and $\zeta$\,, and so
\begin{equation*}
\left|f(c) - f(\zeta)\right|\leqslant\|f'\|_{\infty}\left|c - \zeta\right|\,,
\end{equation*}
where $\|f'\|_{\infty}:={\rm max}\left|f'(x)\right|$\,.
Hence,
\begin{equation*}
\left|\overline{f\,g} - \overline{f}\,\overline{g}\right|\leqslant\|f'\|_{\infty}\left(\int\limits_{-\infty}^{\infty}\left|c - \zeta\right|W(\zeta)\,{\rm d}\zeta\,\right)\overline{g}\,.
\end{equation*}
\end{document}